\documentclass[journal,article,submit,moreauthors,pdftex,10pt,a4paper]{mdpi}

\pdfoutput=1
\preto{\abstractkeywords}{\nolinenumbers}
\firstpage{1}
\articlenumber{x}
\doinum{10.3390/------}
\pubvolume{xx}
\pubyear{2017}
\copyrightyear{2017}
\externaleditor{Academic Editor: name}
\history{Received: date; Accepted: date; Published: date}

\Title{On the uniqueness theorem for pseudo-additive entropies}

\Author{
Petr Jizba $^{1,\ddagger}$* and Jan Korbel $^{1,2\ddagger}$}

\AuthorNames{Petr Jizba, Petr Jizba and Jan Korbel}

\address{%
$^{1}$ \quad FNSPE, Czech Technical University in Prague, B\v{r}ehov\'{a} 7, 115 19, Prague, Czech Republic; p.jizba@fjfi.cvut.cz\\
$^{2}$ \quad Section for Science of Complex Systems, Medical University of Vienna, Spitalgasse 23, 1090 Vienna, Austria; korbeja2@fjfi.cvut.cz}

\corres{Correspondence: p.jizbal@fjfi.cvut.cz; Tel.: +420-224-358-295}

\secondnote{These authors contributed equally to this work.}

\abstract{We discuss  the idea  that  the  Tsallis-type  ($q$-additive)  entropic  chain rule allows  for a wider class of entropic functionals than previously thought. In particular, we point out that the ensuing entropy solutions (e.g., Tsallis entropy)  can be determined uniquely only when one fixes the prescription for handling conditional entropies. Our point is substantiated with the Dar\'{o}tzy's mapping theorem and DeFinetti--Kolmogorov theorem for escort distributions and illustrated with a number of examples.
Connection with Landsberg's classification of non-extensive thermodynamical systems is also briefly discussed.
}
\keyword{pseudo-additive entropy; entropic chain rule; conditional entropy; Dar\'{o}tzy's mapping}

\begin{document}


\section{Introduction}

During the last two decades, the complex dynamical systems undergone an important
conceptual shift. The catalyst was infusion of new ideas from theory of critical phenomena
(scaling laws), (multi)fractals and trees, renormalization group, random matrix theory
and information theory. On the other hand, the usual Boltzmann--Gibbs statistics (BGS)
has proven to be grossly inadequate in this context. While successful in describing stationary
systems characterized by ergodicity or metric transitivity, BGS fails to reproduce
statistical behavior of many real-world systems in biology, astrophysics, geology, and the
economic and social sciences. In recent years the use of a new paradigm known as  generalized
statistics has become very popular in complex systems. The notion ``generalized
statistics'' refers to statistical systems that are described via broad, (semi)heavy-tail distributions.
Examples include generalized hyperbolic distributions, Meixner distributions,
Weibull distributions, and various power-law tail distributions (e.g., Zipf--Pareto, L\'{e}vy
or Tsallis-type distributions). In particular, the statistics associated with power-law tails
accounts for a rich class of phenomena often observed in complex systems ranging from
financial markets, physics and biology to geoscience (see, e.g. Refs.~\cite{Tsallis(2009),Naudts(2011)} and citations therein).

%

Currently there is a number of key techniques for
dealing with generalized-statistics-based complex systems. Among these one can especially
mention; a) innate general statistics approaches such as $q$-deformed thermostatistics of
Tsallis~\cite{Tsallis(2009)}, Superstatistics of Beck et al.~\cite{Beck(2003)} and various generalized non-extensive entropies~\cite{IS(2014),Biro(2015),HTb(2011),HT(2011),Tempesta(2011)};
b) information-theoretic concepts such as information theory of R\'{e}nyi~\cite{JA(2004),JD(2016)},
transfer entropies~\cite{Sch(2000),JKS(2012)}, complexity theory~\cite{Tsallis(2009),Yo(2015)}, or information geometry~\cite{Bercher(2013)}.
Particularly the concept of entropy both in its information-theoretic and combinatorial
disguise plays a central r\^{o}le here.
%
%
%
It is, however, clear that entropy can provide meaningful and technically sound modus operandi only if there is a
good guiding principle behind its formulation. Admittedly, from a mathematical point of view,  the most satisfactory prescriptions are
based on axiomatization of entropy or (and even better) on some operationally viable coding theorem. It is purpose of this paper
to focus on the axiomatic underpinning of entropies with a special emphasize on the axiomatics of the so-called $q$-additive entropies.
We shall show, by extending our previous argument~ \cite{PJ(2017)}, that commonly used axioms for $q$-additive entropies are prone to have multiplicity of distinct solutions with the culprit residing in the way how conditional entropies are handled.
Since the existence of different solutions is intimately related to Dar\'{o}tzy's mapping theorem and Kolmogorov--Nagumo's quasi-linear means,
we study thus generated class of entropic functionals that all satisfy the same $q$-additive entropic chain rule.
We will also briefly touch upon the possibility of yet another mechanism, which is implied by the DeFinetti--Kolmogorov theorem for escort distributions.

The plan of the paper is as follows. After this Introduction, we present in Section~\ref{sec1} a concise overview of some standard (pseudo-)additive entropic rules which yield degenerate solutions for entropic functionals.  This will be important in Section~\ref{sec2}
where the same phenomenon will be observed in the context of entropic chain rules. Despite their higher restrictiveness, the chain rules still lead to degenerate solutions and this fact is present even in the (more sophisticated) $q$-extensive entropic chain rules as demonstrated in Section~\ref{sec3}. In Section~\ref{sec4} we identify the root cause of the degeneracy in the way how conditional entropies are handled.
In particular, we prove that with the help of Dar\'{o}tzy's mapping theorem and Kolmogorov--Nagumo's quasi-linear means one may easily
satisfy the same $q$-extensive entropic chain rule with two completely distinct yet perfectly legitimate entropic functionals.
Finally, Section~\ref{sec5} summarizes our results and discusses further consequences.

\section{Entropic (pseudo-)additivity rules \label{sec1}}

\subsection{Some (pseudo-)additivity rules}

Suppose $X$ and $Y$ are two events, e.g. messages, having $W_X$ and $W_Y$ as possible number of states, respectively.
The two events taken together form a joint event that we denote as $XY$. If $X$ and $Y$ are
independent of each other, every combination of $X$ and $Y$ values is a possible joint event,
and so $W_{XY} = W_X W_Y$. In this case the entropy for a composite
bipartite system $XY$ --- the joint entropy $H(X,Y)$, is additive, i.e.
\begin{eqnarray}
H(X,Y) \ = \ H(X) \ + \ H(Y)\, .
\label{2.1.1.a}
\end{eqnarray}
On the other hand, if the events are not independent (and do not interact), it may be that some combinations
of $X$ and $Y$ are not allowed (i.e., $W_{XY} < W_X W_Y$), so that the joint entropy $H(X,Y)$ may be less than $H(X) + H(Y)$ --- a property
known as the {\em subadditivity} of the entropy.
We will have more to say about this in Section~\ref{sec4}.


In practice, the entropy behavior in a number of independent bipartite systems cannot be easily grasped via simple additivity prescription (\ref{2.1.1.a}). The reasons behind might be the fact that some marginal states (e.g., surface states) are neglected or that the long-range correlations are invalidating the assumption of independence and the concept of independence is used only as a convenient approximation. Whatever the reason,  the modus operandi in these cases is to resort to some simple one-parameter deformation of the additivity rule  that grasps in one way or another the non-additive contributions. Such deformations are known as {\em pseudo-additive} entropic forms. Most commonly used versions are; Tsallis-type additivity~\cite{Tsallis(2009)}
\begin{eqnarray}
H(X,Y) \ = \ H(X) \ + \ H(Y) \ + \ (1-q)H(X)H(Y)\, ,
\end{eqnarray}
Landsberg-type additivity~\cite{Landsberg(1999)}
\begin{eqnarray}
H(X,Y) \ = \ H(X) \ + \ H(Y) \ + \ (q-1)H(X)H(Y)\, ,
\end{eqnarray}
(AdS/CFT) $\delta$-additivity~\cite{Tsallis(2009),vos(2015)}
\begin{eqnarray}
H(X,Y)^{1/\delta} \ = \ H(X)^{1/\delta} \ + \ H(Y)^{1/\delta}\, ,
\label{2.1.4.a}
\end{eqnarray}
Masi--Czachor-type supra-additivity~\cite{Masi(2005)}, etc.

Let us point out that it might happen that the above (pseudo-)additivity rules hold also for more generic joint systems for which $W_{XY} \neq W_X W_Y$. In such cases one speaks about (pseudo-)extensivity~\cite{HT(2011)}. Since  the (pseudo-)extensivity is more a trade of the actual interaction/correlation between $X$ and $Y$ rather than entropy itself we shall not dwell on this issue here. Let us just mention that the actual form of non-extensivity and the value of parameters can be often connected to some physical phenomena \cite{Biro(2015),Korbel(2017)}. In Section~\ref{sec2} we will see that,  from a strictly mathematical standpoint, it is logically more satisfactory to deal with dependent but non-interacting systems. In such cases above (pseudo-)additivity rules will be replaced with more restrictive {\em entropy chain rules}.

\subsection{Degeneracy in solutions }

It is well known that there is a number of ``logically consistent'' but form-inequivalent entropic functionals satisfying the simple additivity rule (\ref{2.1.1.a}). Examples are provided by Shannon entropy (and ensuing Gibbs and von Neumann entropies) or R\'{e}nyi entropy.
In this connection, we should emphasize perhaps a less known fact that a similar situation holds also for pseudo-additivity. For instance,
by assuming that with $X$ is associated probability $\mathcal{P}_X = \{ p_i \}_{i=1}^n$,
then the Tsallis-type additivity rule is not only satisfied with Tsallis entropy
\begin{eqnarray}
H(X) \  \equiv \ H(\mathcal{P}_X) \ = \ \frac{\sum_{i=1}^n p_i^{q} - 1}{1-q}\, ,
\end{eqnarray}
as one would expect, but also with Landsberg's entropy~\cite{Landsberg(1999)}
\begin{eqnarray}
H(X) \  = \  \frac{\left(\sum_{i=1}^n p_i^{2-q}\right)^{-1} -1}{1-q}\, ,
\end{eqnarray}
or with Behara--Chawla $\gamma$-entropy~\cite{Behara}
\begin{eqnarray}
H(X) \  = \  \frac{1- \left(\sum_{i=1}^n p_i^{1/\gamma}\right)^{\gamma}}{1-2^{\gamma-1}}\, ,  \,\,\,\,\, q \ = \ 2- 2^{\gamma-1}\, .
\end{eqnarray}

On the same footing, the (AdS/CFT) $\delta$-additivity is satisfied with a class of the so-called Segal entropies~\cite{segal(1978)}.

\subsection{First bite on uniqueness}

At this stage there are two points that should be noted. First, above (pseudo-)additive forms (\ref{2.1.1.a})-(\ref{2.1.4.a}) are true only for independent events. Second, one could entertain the idea that when the entropic {\em chain rules} (which employ also the conditional entropy $H(Y|X)$)  are used instead, then their higher restrictiveness could remove the unwanted degeneracy and lead to a single unique solution. This might seem as a good strategy particularly because the entropic chain rule with ensuing $H(Y|X)$ is of practical relevance in a number of ares. In this connection one can mention applications in;
\begin{enumerate}[leftmargin=10mm,labelsep=4.9mm]
\item {\em information theory} where it describes input-output information of (quantum) communication channel and enters in data processing inequality~\cite{campbell(1965)},
\item {\em non-equilibrium and complex dynamics}  where it enters in May--Wigner criterion for the stability of  dynamical systems and helps to estimate the connectivity of the network of system exchanges~\cite{hastings(1982)},
\item {\em time series and data analysis} where it describes  transfer entropies in bivariate time series and degree of synchronization between two signals~\cite{JKS(2012)} .
\end{enumerate}
%
%
%
%
In the following two sections we will see what the entropic chain rules and their pseudo-additive extensions can really say about the uniqueness of entropic functionals.

\section{Entropic chain rule \label{sec2}}

\subsection{Additive entropic chain rule --- some fundamentals}

Let us consider two generally dependent events. Elementary (Shannon-type) entropic chain rule for ensuing random variables $X$ and $Y$  reads
\begin{eqnarray}
H(X,Y) \ = \ H(X) \ + \  H(Y|X)\, ,
\label{3.1.a}
\end{eqnarray}
where $H(X)$  is a single-random variable entropy,  $H(X,Y)$ is joint entropy,
and $H(Y|X)$ represents the related conditional entropy of $Y$ given $X$. The meaning of the chain rule (\ref{3.1.a}) is depicted in Figure~1.
%
%
\begin{figure}[H]
\centering
\includegraphics[width=4.7cm]{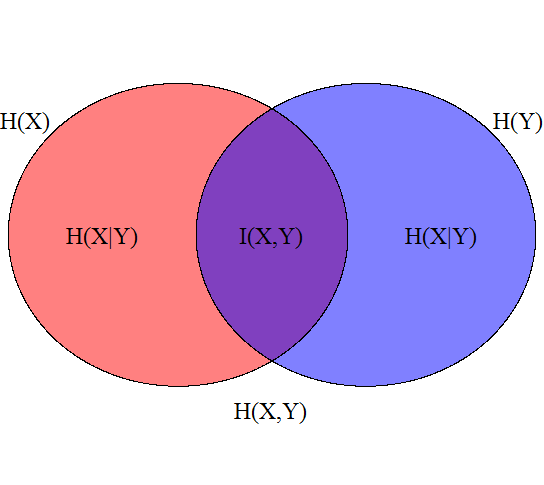}
\caption{Entropy Venn diagram for two random variables. The symbol $I(X,Y)$ denotes the {\em mutual information}, i.e., $I(X,Y) = H(X) + H(Y) - H(X,Y)$.}
\end{figure}
By induction one can generalize the previous relation (\ref{3.1.a}) to
\begin{eqnarray}
H(X_1,X_2, \ldots, X_n) \! &=& \! H(X_1)  \ + \  H(X_2|X_1)  \ + \  H(X_3|X_2, X_1) \  + \  H(X_4|X_3, X_2, X_1) \ + \ \ldots \nonumber \\[2mm]
&=& \! \sum_{i=1}^n H(X_i|X_{i-1}, X_{i-2}, \ldots, X_1)\, .
\label{3.2.ab}
\end{eqnarray}
At this stage we can ask ourselves a following question. By imposing simple consistency conditions (i.e., Kolmogorov axioms 1-3, see, e.g.~\cite{HTb(2011)}), such as:
\begin{enumerate}[leftmargin=10mm,labelsep=4.9mm]
  \item \emph{continuity}, i.e., when $H(\mathcal{P}_X)$ is a continuous function of all arguments,
  \item \emph{maximality}, i.e., when for given $n$ is $H(\mathcal{P}_X)$ maximal for $\mathcal{P}_X = \left({1}/{n}, \dots, {1}/{n}\right)$ and
  \item \emph{expansibility}, i.e., when $H(p_1,\dots,p_n,0) = H(p_1,\dots,p_n)$,
\end{enumerate}
%
%
to what extend is  $H(X)$ (with the above chain rule) unique?
%
%
%
It might perhaps come as a surprise that the uniqueness is not guarantied, at least not unless one prescribes how to handle \emph{conditional}  entropies.

To understand this point better, let us state the basic defining properties of $H(X|Y)$:
\begin{enumerate}[leftmargin=10mm,labelsep=4.9mm]
  \item $H(Y|X)\ = \ 0$ iff $Y$ is completely determined by $X$, e.g. $H(g(X)|X) = 0$ ($g$ is some function),
  \item $H(Y|X) \ = \ H(X|Y) \ - \ H(X) \ + \ H(Y)$  \emph{entropic Bayes' rule},
  \item $H(X|Y) \ \leq \ H(X)$ \emph{``second law od thermodynamics''},
  \item $H(X|Y) \ = \ H(X)$ if $Y$ and $X$ are independent (sometimes ``if and only if''  is required).
\end{enumerate}
%
These properties still allow for some additional flexibility. Examples might be provided with Shannon's solution and
R\'{e}nyi's solution which both share the same chain condition (\ref{3.2.ab}) and the above defining conditions for $H(X|Y)$. In particular,
for Shannon's entropy the conditional entropy of  $X$ give $Y$ is constructed in two step
\begin{eqnarray}
&&H(X|Y = y) \ = \ - \sum_{x \in X} p(x|y) \log_2 p(x|y)\, ,\nonumber \\[2mm]
&&H(X|Y) \ = \ \sum_{y \in Y}p(y) H(X|Y = y) \ = \ -\!\! \sum_{x\in X; y\in Y} p(x,y)
\log_2 p(x|y)\, ,
\end{eqnarray}
i.e., $H(X|Y)$  is defined via linear mean. In contrast, in the case of R\'{e}nyi's entropy the conditional entropy of $X$ given $Y$ is defined by means of quasi-linear Kologorov--Nagumo (KN) mean~\cite{Nagumo(1930),Aczel} in the following two-step sequence~\cite{Renyi}
\begin{eqnarray}
&&I_q(X|Y = y) \ = \ \frac{1}{1-q}\log_2\sum_{x\in X} p^q(x|y)\, ,\nonumber \\[2mm]
&&I_q(X|Y) \ = \ \phi^{-1}\left[\sum_{y \in Y}\rho_q(y) \phi\left(I_q(X|Y = y)\right)\right] \ = \ \frac{1}{1-q}\log_2 \frac{\sum_{x\in X; y\in Y} p^q(x,y)}{\sum_{y\in Y} p^q(y)}\, .
\label{11.abcd}
\end{eqnarray}
Here, the KN function $\phi$ reads~\cite{JA(2004),Aczel}
\begin{eqnarray}
\phi(x) \ = \ \exp[(1-q)x]\, .
\end{eqnarray}
Particularly noteworthy is the appearance of so-called {\em escort distribution}~\cite{Beck,Renyi}:
\begin{eqnarray}
\rho_q(y) \ = \ \frac{p^q(y)}{\sum_{y\in Y} p^q(y)}\, ,
\end{eqnarray}
in Eq.~(\ref{11.abcd}). Interestingly, a need for $\rho_q(y)$ in the definition of $I_q(X|Y)$ was already
observed by A.~R\'{e}nyi in 1960's~\cite{Renyi} --- some 30 years before the escort distribution was officially introduced in Ref.~\cite{Beck}.

\section{$q$-extensive entropic chain rule \label{sec3}}
\subsection{Second bite on uniqueness}

In analogy with the preceding section one defines the $q$-extensive entropic chain rule for two random generally dependent variables $X$  and $Y$  as
\begin{eqnarray}
H(X,Y) \ = \ H(X) \ + \  H(Y|X) \ + \ (1-q) H(X) H(Y|X)\, .
\end{eqnarray}
By induction one can generalize this to an $n$-partite system as
\begin{eqnarray}
H(X_1,X_2, \ldots, X_n)  &=&  \sum_{i=1}^n H(X_i|X_{i-1},  \ldots, X_1) \nonumber \\[1mm]
&+& (1-q) \sum_{i<j} H(X_i|X_{i-1},  \ldots, X_1) H(X_j|X_{j-1},  \ldots, X_1) \nonumber
\\[1mm]
&+& (1-q)^2 \sum_{i<j<k} H(X_i|X_{i-1},  \ldots, X_1) H(X_j|X_{j-1}, \ldots, X_1) H(X_k|X_{k-1}, \ldots, X_1) \nonumber
\\
&+&  \ldots \; +  \ (1-q)^{n-1} \prod_{i=1}^n H(X_i|X_{i-1}, X_{i-2}, \ldots, X_1)\, .
\end{eqnarray}
We can ask again a similar question as before, namely to what extend is  $H(X)\equiv H(\mathcal{P}_X)$ (subject to above chain rule) unique.

As before, we should not be surprised to learn that the uniqueness is not guarantied unless $H(X|Y)$ is specified. Indeed, the $q$-extensive entropic chain rule together with consistency conditions
\begin{enumerate}[leftmargin=10mm,labelsep=4.9mm]
\item $H(Y|X) \ = \ 0$ iff $Y$ is completely determined by $X$, e.g. $H(g(X)|X) = 0$,
\item $\Phi(X,q)H(Y|X) \ = \  H(Y) \ - \ H(X) \ + \ \Phi(Y,q)H(X|Y)$  \emph{$q$-entropic Bayes' rule}\footnote{The explicit form of the function $\Phi(X,q)$ depends on the type of the $q$ deformation. For instance, for $q$ extensivity we have $\Phi(X,q) = 1 + (1-q) H(X)$.},
\item $H(X|Y) \ \leq \ H(X)$ \emph{``second law od thermodynamics''},
\item $H(X|Y) \ = \ H(X)$ if $Y$ and $X$ are independent,
\end{enumerate}
are not enough to fix the unique form of $H(X)$. Examples include
%
%
%
%
%
a) Tsallis' entropy where the conditional entropy of $X$ given  $Y$ is defined as~\cite{Abe}
\begin{eqnarray}
  &&S_q(X|Y = y) \ = \ \frac{1}{1-q}\left[\sum_{x \in X} p(x|y)^q - 1 \right]\, ,\nonumber\\
  &&S_q(X|Y) \ = \ \sum_{y \in Y}\rho_q(y) S_q(X|Y = y)\, ,
  \end{eqnarray}
i.e., $S_q(A|B)$ is defined via linear (escort) mean,
b) Frank--Daffertshofer entropy~\cite{FD} where the conditional entropy is defined as
\begin{eqnarray}
  &&S_{FD}(X|Y = y) \ = \ \frac{1}{1-q}\left[\left(\sum_{x \in X} p(x|y)^r\right)^{\frac{1-q}{1-r}} - 1\right]\, ,\nonumber\\
  &&S_{FD}(X|Y) \ = \ \phi^{-1}\left[\sum_{y \in Y}\rho_q(y) \phi\left(S_{FD}(X|Y = y)\right)\right]\, ,
  \label{SM13}
\end{eqnarray}
with the KN function $\phi(x) = \log_{r} e_q^x$, i.e., $S_{FD}(X|Y)$ is obtained via quasi-linear KN mean,
c) Sharma--Mittal entropy~\cite{SM} where the conditional entropy is given by
\begin{eqnarray}
&&S_{SM}(X|Y = y) \ = \ \frac{1}{\delta}\left[\left(\sum_{x \in X} p(x|y)^r\right)^{\frac{1-q}{1-r}} - 1\right]\, ,\nonumber\\
&&S_{SM}(X|Y) \ = \ \phi^{-1}\left[\sum_{y \in Y}\rho_q(y) \phi\left(S_{SM}(X|Y = y)\right)\right]\, .
\label{SM14}
\end{eqnarray}
Here $\delta = 2^{1-q} - 1$ and  $\phi(x) = \log_{r} e_{\delta  - 1}^x$, i.e.  $S_{SM}(X|Y) $ is again defined via quasi-linear KN mean.
%

\section{Two simple theorems \label{sec4}}


Let us now state two simple theorems that are pertinent to the discussion of $q$-extensive entropic rules.

\begin{Theorem}
{\bf  \em  Dar\'{o}tzy's mapping:} is a monotonic mapping $h_{\gamma}: \mathbb{R} \rightarrow \mathbb{R}$ such that
\begin{eqnarray*}
h_{\gamma}(x \ + \ y) \ = \  h_{\gamma}(x) \ + \ h_{\gamma}(y) \ + \ \gamma h_{\gamma}(x)h_{\gamma}(y) \ \equiv \ h_{\gamma}(x) \oplus_{\gamma} h_{\gamma}(y)\, ,
\end{eqnarray*}
$h_{\gamma}$ is parameterized by $a, \lambda$ and $\gamma$ so that
\begin{eqnarray*}
h_{\gamma}(x) \ = \ \left\{
                      \begin{array}{ll}
                        ax, & a > 0~ \mbox{for} ~\lambda = 0 \\[2mm]
                        (2^{\lambda x} -1)/{\gamma}, & \lambda\gamma >0 ~\mbox{for}~ \lambda \neq 0\, .
                      \end{array}
                    \right.
\end{eqnarray*}\\[3mm]
{\em \bf Statement:} Let us have additive chain rule with $H(X|Y)$ defined with KN function $\phi(x)$. Dar\'{o}tzy's mapping then generates $q$-extensivity (provided we set $\gamma = 1-q$) where new $H(X|Y)$ is defined via KN function $\phi\circ h^{-1}_{\gamma}$.
\end{Theorem}

\begin{proof}[Proof of Theorem 1]
\begin{eqnarray*}
H(X,Y) \ = \  H(X) \ + \ H(Y|X) \ \Rightarrow \ h_{\gamma}(H(X,Y)) \ = \   h_{\gamma}(H(X)) \ \oplus_{\gamma} \ h_{\gamma}(H(X|Y))\, ,
\end{eqnarray*}
where
\begin{eqnarray*}
h_{\gamma}(H(X|Y))
\ = \
h_{\gamma}\circ \phi^{-1} \left[\sum_{y \in B}\rho_q(y) \ \! \phi \circ h^{-1}_{\gamma}(h_{\gamma}(H(X|Y=y)))\right].
\end{eqnarray*}
The actual values of $a$ and $\lambda$ are immaterial for the $q$-extensivity as the first line of the proof does not depend on them. The second line shows that the ensuing $q$-extensive entropy represents a two-parametric functional (apart from parameters originating from the KN function $\phi$).
\end{proof}
So, the core idea of this theorem is that if there exist two {\em different} entropy functionals for a given {\em additive} chain rule with two different KN-related conditional entropies, then one can use Dar\'{o}tzy's mapping to map these two solutions to another two solutions that solve the $q$-extensive chain rule. A simple examples of this mechanism were provided by the Frank--Daffertshofer and Sharma--Mittal entropies (\ref{SM13})-(\ref{SM14}) with  Dar\'{o}tzy's mappings $h_{\delta}(x) = (2^{(1-q)x}-1)/\delta$ applied to R\'{e}nyi entropy $I_r$ with $\delta = 2^{1-q} -1$ and $\delta = 1-q$, respectively. For some further details on this issue see, e.g. Ref.~\cite{IS(2014)}.



The second theorem highlights some tricky points related to {\em joint} escort distributions.

\begin{Theorem}
{\bf \em DeFinetti--Kolmogorov relation and escort distributions:}
when working with escort distributions the joint distributions derived from the original joint distributions $\{r_{kl}\}$ do not satisfy DeFinetti--Kolmogorov relation, i.e. analogue of $r_{kl} = p_l r_{k|l}$ (here $\{p_l\}$ and $\{r_{k|l}\}$ are  marginal and conditional distributions, respectively).
\end{Theorem}


\begin{proof}[Proof of Theorem 2]
By using the standard DeFinetti--Kolmogorov relation for original distributions, i.e. $r_{kl} = p_l r_{k|l}$, we may write the following chain of relations:
\begin{eqnarray*}
r_{kl} \ = \ p_l r_{k|l} \ \Leftrightarrow \ r_{kl}^q \ = \ p_l^q r_{k|l}^q \ \Leftrightarrow \ R(q)_{kl}&=& \frac{r_{kl}^q}{\sum_{mn}r_{mn}^q} \ = \ \frac{p_l^q r_{k|l}^q}{\sum_{mn}r_{mn}^q}\nonumber \\
&=& \frac{p_l^q r_{k|l}^q}{\sum_{n}(p_n^q \sum_m r_{m|n}^q)} \ \neq \ \frac{p_l^q}{\sum_n p_n^q}\frac{r_{k|l}^q}{\sum_{m} r^q_{m|l}} \ \equiv \ \tilde{R}(q)_{kl}\, .
\end{eqnarray*}
Here $\tilde{R}(q)_{kl}$ is the correct would-be joint escort distribution. Note that $\tilde{R}(q)_{kl}$ is not the escort of $r_{kl}$ and  $\tilde{R}(q)_{kl} = R(q)_{kl}$ iff events are independent~\cite{PJ(2017)}.
\end{proof}
An important upshot  of this theorem is that some entropies (particularly those born out of escort distributions) might satisfy  $q$-additivity but not necessarily $q$-extensive entropic chain rule. The reason is that the more restrictive chain rule is sensitive to the failure of the DeFinetti--Kolmogorov relation for joint escort distributions. Typical example is provided by the JA entropy~\cite{PJ(2017),Jizba(2016),Cankaya(2017)}
\begin{eqnarray*}
S_{JA,q}(X) \ = \ \frac{1}{1-q}\left(2^{-(1-q)\sum_k \rho_k(q) \log_2 p_k}  -1 \right) \ = \ h_{1-q}\left(-\sum_k \rho_k(q) \log_2 p_k\right),
\end{eqnarray*}
where $h_{1-q}(x)$ is  the KN function  $h_{1-q}(x) =  (2^{(1-q)x}-1)/(1-q)$. Because $-\sum_k \rho_k(q) \log p_k$ satisfies the simple additivity rule for {\em independent} events, Dar\'{o}tzy's mapping ensures that $S_{JA,q}(X)$ satisfies the $q$-additivity.  On the other hand,  $-\sum_k \rho_k(q) \log p_k$ does not satisfy the simple chain rule due to failure of the DeFinetti--Kolmogorov relation for joint events. Consequently, $S_{JA,q}(X)$ is not $q$-extensive.

%


In this connection we might recall Landsberg's classification of non-extensive thermodynamics.
In 90's P.T.~Landsberg classified  types of thermodynamics according to all possible functional properties of the entropy~\cite{Landsberg(1999)}. By defining entropy classes
\begin{enumerate}[leftmargin=10mm,labelsep=4.9mm]
\item {\em Superadditivity} - {~\bf $S$}:~~  $H(\mathcal{P}_{X+Y}) \ \geq \ H(\mathcal{P}_{X}) \ + \ H(\mathcal{P}_{Y})$,
\item {\em Homogeneity} - {~\bf $H$}:~~  $H(\mathcal{P}_{\lambda X}) \ = \ \lambda H(\mathcal{P}_{X})$,
\item {\em Concavity} - {~\bf $C$}:~~ $H(\mathcal{P}_{\lambda X + (1-\lambda)Y}) \ \geq \ \lambda H(\mathcal{P}_X) \ + \ (1-\lambda) H(\mathcal{P}_Y)$.
\end{enumerate}
Landsberg concluded that there are only 6 logically consistent thermodynamic classes: {\bf $SHC$}, {\bf $S{\bar{H}}C$},  {\bf $S\bar{H}\bar{C}$},  {\bf $\bar{S}\bar{H}C$}, {\bf $\bar{S}H\bar{C}$}, {\bf $\bar{S}\bar{H}\bar{C}$}. Types {\bf $SH\bar{C}$} are {\bf $\bar{S}HC$} are logically impossible~\cite{Landsberg(1999)}, see Figure~2.
\begin{figure}[H]
\centering
\includegraphics[width=4.7cm]{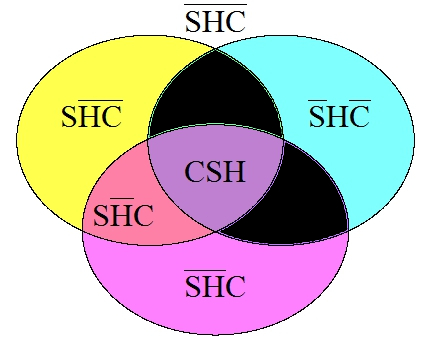}
\caption{Venn diagram for Landsberg's thermodynamic types.}
\end{figure}
This classification is, however, based entirely on entropies for independent events.
Since entropic (pseudo-)additivity rule and chain rule for a given system  do not need to follow the same pattern (e.g., they do not need to be both $q$-additive) the entropy class might change when conditional probabilities are included.
For instance, according to Landsberg $S_{JA,q}(X)$ should for $q<1$,  belong to {\bf ${S}\bar{H} C$} class, while the corresponding chain-rule
generalization allows for flipping between $S$ and $\bar{S}$, see~\cite{PJ(2017)}.

\section{Conclusions \label{sec5}}

In this paper we have demonstrated that any given $q$-extensive {\em entropic chain} rule allows for a wider class of entropic functionals than previously thought. This degeneracy in solutions plagues also $q$-additive entropic rules but one would expect that the higher restrictiveness of the $q$-extensive rule could perhaps remove such a degeneracy. This is not the case. The culprit behind is in the way how $H(X|Y)$ is handled. There is a flexibility in the definition of permissible $H(X|Y)$ by using  quasi-linear (or KN) means. These results beg for a question; what is a typical trademark of non-extensive statistics; a)  pseudo-additivity (as indicated by Landsberg's classification), b)  pseudo-extensivity (i.e., chain rule with conditional entropies) or c)  power-law-type entropy maximizers.
%

%
%
%
%
%

\vspace{6pt}


\acknowledgments{Both P.J. and J.K.  were  supported  by the Czech  Science  Foundation Grant No. 17-33812L. J.K. was also supported by the Austrian Science Fund, Grant No. I 3073-N32.}


\conflictsofinterest{The author declares no conflict of interest.}

\abbreviations{The following abbreviations are used in this manuscript:\\
\noindent
\begin{tabular}{@{}ll}
KN & Kolmogoron--Nagumo\\
\end{tabular}}

%


\bibliographystyle{plain}
\reftitle{References}




\end{document}